\documentclass{webofc}
\usepackage[varg]{txfonts}

\newcommand{\ds}{\displaystyle}
\newcommand{\vev}[1]{\langle#1\rangle}
\newcommand{\mat}{\left ( \begin{array}}
\newcommand{\emat}{\end{array} \right )}
\newcommand{\vect}{\left ( \begin{array}{c}}
\newcommand{\evect}{\end{array} \right )}

\begin{document}
\title{Dense quark matter with chiral and isospin imbalance: NJL-model consideration
}

\author{\firstname{Tamaz} \lastname{Khunjua}\inst{1}\fnsep\thanks{\email{tg.khunjua@physics.msu.ru}} \and
        \firstname{Konstantin} \lastname{Klimenko}\inst{2}\fnsep\thanks{\email{konstantin.klimenko@ihep.ru}} \and
        \firstname{Roman} \lastname{Zhokhov}\inst{2,3}\fnsep\thanks{\email{zhokhovr@gmail.com}}
      }

\institute{Faculty of Physics, M.V. Lomonosov Moscow State University, Moscow
\and
           Logunov Institute for High Energy Physics, NRC "Kurchatov Institute", Protvino, Moscow Region
\and
           Pushkov Institute of Terrestrial Magnetism, Ionosphere and Radiowave Propagation (IZMIRAN), Troitsk, Moscow
          }

\abstract{
Isospin asymmetry is the well-known property of dense quark matter, which exists in the compact stars and is produced in heavy ion collisions. On the other hand, the chiral imbalance between left- and right- handed quarks is another highly anticipated phenomenon that could occur in the dense quark matter. To investigate quark matter under these conditions, we take into account baryon -- $\mu_B$, isospin -- $\mu_I$ and chiral isospin -- $\mu_{I5}$ chemical potentials and study QCD phase portrait using NJL$_4$ model generalized to two massive quarks that could condense into the pion condensation. We have shown that the chiral isospin chemical potential $\mu_{I5}$ generates pion condensation in isospin asymmetric quark matter. Also, we have investigated discrete symmetry (duality) between chiral and pion condensates in the case of massless quarks, which stay relatively instructive even if the quarks have bare mass. To describe hot dense quark matter, in addition to the above-mentioned chemical potentials, we introduce non-zero temperatures into consideration.
}
\maketitle
\section{Introduction}
\label{intro}

The fundamental theory of the dense quark matter is quantum chromodynamics (QCD) which is a gauge field theory associated with $SU(3)$ group, where gauge bosons (gluons) play the role of interaction carriers and the spinor representation of the group is associated with quarks. The main method of QCD analysis is the perturbative technique
in
coupling
 constant. However, it is not always possible to use this technique, as QCD calculations can be too complex or the coupling constant can be too large due to asymptotic freedom. The situation becomes even more difficult for the quark matter with non-zero baryon chemical potential (i.e. dense quark matter). In these cases, non-perturbative methods, such as effective theories or lattice calculations are usually used.

Lattice calculations are very useful for description of the region of zero density and high temperature. However, the so-called sign problem still presents insurmountable difficulties for lattice calculations in the non-zero density region. Nevertheless, at this moment, effective models
are
 the best tool for investigating dense quark matter. One of the most widely used effective model
%
%at this time
today
 is Nambu--Jona-Lazinio (NJL) model. It was originally formulated to describe nucleon mass creation via spontaneous breaking of chiral symmetry in analogy to classical superconductivity and was based on nucleons, pions, and scalar $\sigma$-mesons \cite{nambu1961}. Later, it was reformulated for quarks and
 it was shown
that light quarks acquire mass as a result of spontaneous breaking of chiral symmetry.

Due to cooling effects and electrical neutrality dense baryonic matter in compact stars
%
%obeys
possesses
an isospin asymmetry, i.e. the densities of up- and down quarks are different (it
is characterized
 by isospin chemical potential $\mu_I$). In experiments on heavy-ion collisions, we also have to deal with dense baryonic matter which has an evident isospin asymmetry because of different neutron and proton contents of initial ions. In early
%
%70-th
70-s
%seventies
%
 Sawyer \cite{sawyer1972condensed} and independently Migdal \cite{migdal1973phase} have shown that there might be phase transition from pure neutron matter to mixed hadron matter with protons, neutrons and $\pi^0$-pions at superdense matter in the compact stars. Later, using the chiral perturbation theory ($\chi$PT), it was shown that there is a threshold $\mu_I^c=m_\pi\approx 140 \,{\rm MeV}$ of a second order phase transition to the charged pion condensation phase \cite{Son2000xc,Son2001,Splittorff:2000mm,Loewe2002tw}. This result was ultimately
proved
  in the framework of NJL model \cite{PhysRevD.69.096004,PhysRevD.70.054013, EK2006, AndersenNJL} (including (1+1)-dimensional version of the NJL model \cite{Ebert2009ty,Adhikari2016vuu}) and lattice simulations \cite{Gupta,PhysRevD.66.014508, Brandt2017oyy}.
Nevertheless,
 the whole picture is still a matter of debate.

The main question is pion condensation existence in the real world and
its
 influence on the observable values
%phenomena
.
Indeed there are a lot of causes that could promote or suppress the PC phase. For example, in the framework of NJL model the finite-size effects or spatial
inhomogeneity
 of the condensates could promote the PC phase \cite{Ebert2011tt,Gubina2012wp,Khunjua2017khh}. On the other hand, PC phase could be suppressed in $\beta$-equilibrium and
its
 existence strongly depends on bare quark mass, temperature and model parameters \cite{EK2006n,AndersenNJL}.

Recently, it has been shown that chiral imbalance
promotes
 PC phase in the framework of the NJL model \cite{Khunjua2017mkc}.

 Imbalance between left-handed and right-handed quarks is highly anticipated phenomena that could occur both in compact stars and heavy ion collisions. This effect could stem from nontrivial interplay of axial anomaly and the topology of gluon configurations. Also, there is another mechanism -- chiral separation effect (CSE) which is realized in strong magnetic field and leads to chiral asymmetry. In the 2-flavor case CSE could promote both non-zero chiral density $n_5$ and non-zero isotopic chiral density $n_{I5}$.

In the present work we investigate the phase portrait and pion condensation phenomenon of the isotopic and chiral isotopic imbalanced dense quark matter ($\mu_B\neq0;\,\mu_I\neq0;\,\mu_{I5}\neq0$) with non-zero bare quark mass and temperature using NJL-like model, i.e. we generalize the work \cite{Khunjua2017mkc} to the case of
 the non-zero bare quark mass and finite temperature.

\section{The model and its thermodynamic potential}
\label{sec-1}
It is well known that in the framework of effective four-fermion field theories dense and isotopically asymmetric quark matter, composed of $u$ and $d$ quarks, can be described by the following (3+1)-dimensional NJL Lagrangian
\begin{eqnarray}
&&  L=\bar q\Big [\gamma^\nu\mathrm{i}\partial_\nu -m_0
+\frac{\mu_B}{3}\gamma^0+\frac{\mu_I}2 \tau_3\gamma^0\Big ]q+ \frac
{G}{N_c}\Big [(\bar qq)^2+(\bar q\mathrm{i}\gamma^5\vec\tau q)^2 \Big
]. \label{1}
\end{eqnarray}
Here $q$ is a flavor doublet, $q=(q_u,q_d)^T$, where $q_u$ and $q_d$ are four-component Dirac spinors as well as color $N_c$-plets of the $u$ and $d$ quark fields, respectively (the summation in (\ref{1}) over flavor, color, and spinor indices
are
 implied); $\tau_k$ ($k=1,2,3$) are Pauli matrices; $m_0$ is the bare quark mass (for simplicity, we assume that $u$ and $d$ quarks have the same mass); $\mu_B$ and $\mu_I$ are chemical potentials which are
introduced in order to study quark matter with nonzero baryon and isospin densities, respectively.

The symmetries of the Lagrangian (\ref{1})
depend
 essentially on
whether
 the bare quark mass $m_0$ and chemical potentials take zero or nonzero values. For example, in the most particular case, when $m_0=\mu_{I}=0$ the Lagrangian (\ref{1}) is invariant under transformations from chiral $SU(2)_{L}\times SU(2)_{R}$ group, which is also inherent in 2-flavor QCD in the chiral limit. This symmetry is reduced to $U_B(1)\times U_{I_3}(1)\times U_{AI_3}(1)$ group if all chemical potentials are nonzero, and $m_0=0$.

As a result, we
can
 see that in the chiral limit ($m_0=0$) the quantities $\hat n_B\equiv\bar q\gamma^0q/3$, $\hat n_I\equiv\bar q\gamma^0\tau^3 q/2$ and $\hat n_{I5}=\bar q\gamma^0\gamma^5\tau^3 q/2$ are the densities of conserved baryon, isospin and chiral isospin charges, respectively, of the system (1). Introducing the particle density operators for $u$ and $d$ quarks, $\hat n_u\equiv q_u\gamma^0q_u$ and $\hat n_d\equiv q_d\gamma^0q_d$, we have
\begin{eqnarray}
\hat n_B=\frac 13\left (\hat n_u+\hat n_d\right ),~~\hat n_I=\frac 12\left (\hat n_u-\hat n_d\right ).
\label{2003}
\end{eqnarray}
One can also introduce the particle density operators $\hat n_{fR}$ and  $\hat n_{fL}$ for right- and left-handed quarks of each flavor $f=u,d$. In this case the density of chiral isospin charge looks like
\begin{eqnarray}
\hat n_{I5}=\frac 12\left (\hat n_{uR}-\hat n_{uL}-\hat n_{dR}+\hat n_{dL}\right )=\frac 12\left (\hat n_{u5}-\hat n_{d5}\right ),
\label{2004}
\end{eqnarray}
where the quantity $\hat n_{f5}\equiv \hat n_{fR}-\hat n_{fL}$ is usually called the density of  chiral charge for the quark flavor $f=u,d$. There is a possibility of the appearance of a nonzero chiral isotopic density in quark matter inside neutron stars. Its appearance can be explained on the basis of the chiral separation effect in the presence of a strong magnetic field in a dense baryonic medium.

However, at the physical point ($m_0\neq0$) the symmetry of the Lagrangian (1) under transformations from axial isotopic group $U_{AI_3}(1)$ is explicitly broken. So in the most general case with $m_0\ne 0$, $\mu\ne 0$ and $\mu_I\ne 0$ the initial model (\ref{1}) is invariant under $U_B(1)\times U_{I_3}(1)$ group.

The ground state (the state of thermodynamic equilibrium) of quark matter with $n_B\ne 0$ and $n_I\ne 0$, where $n_B\equiv\vev{\hat n_B}$, $n_I\equiv\vev{\hat n_I}$,\footnote{The notation $\vev{\hat O}$ means the ground state expectation value of the operator $\hat O$.} both at zero and nonzero values of $m_0$ has been investigated in the framework of the NJL model (1), e.g., in Refs. \cite{EK2006,EK2006n,AndersenNJL}. However, the fact that quark matter may have a nonzero chiral isotopic charge was ignored in those papers. Recently, this gap in researches was filled in the paper \cite{Khunjua2017mkc}, where we have studied the properties of equilibrium quark matter at $n_B\ne 0$, $n_I\ne 0$ as well as at nonzero chiral isospin charge density $n_{I5}\equiv\vev{\hat n_{I5}}\ne 0$ in the framework of the massless (3+1)-dimensional two-flavor NJL model (temperature $T$ was taken to be zero in Ref. \cite{Khunjua2017mkc}). In contrast to this, in the present paper we consider the properties of a more realistic quark matter, i.e. at $m_0\ne 0$ and $T\ne 0$, for which all densities $n_B$, $n_I$ and $n_{I5}$ are also nonzero. The solution of this problem can be most conveniently carried out in terms of chemical potentials $\mu_B$, $\mu_I$ and $\mu_{I5}$, which are the quantities, thermodynamically conjugated to corresponding charge densities $\hat n_B$, $\hat n_I$ and $\hat n_{I5}$. Therefore, when solving this problem, one can rely on the Lagrangian of the form\footnote{Generally speaking, in this case the chiral isospin charge is no more a conserved quantity of our system. Therefore, $\mu_{I5}$ is not conjugated to a strictly conserved charge. However, denoting by $\tau$ the typical time scale in which all chirality changing processes take place, one can treat $\mu_{I5}$ as the chemical potential that describes a system in thermodynamic equilibrium with a fixed value of $n_{I5}$ on a time scale much larger than $\tau$.}
\begin{eqnarray}
  \bar L&=&L+\mu_{I5}\hat n_{I5}\nonumber\\
&=&\bar q\Big [\gamma^\nu\mathrm{i}\partial_\nu -m_0
+\frac{\mu_B}{3}\gamma^0+\frac{\mu_I}2 \tau_3\gamma^0+\frac{\mu_{I5}}2 \tau_3\gamma^0\gamma^5\Big ]q+ \frac
{G}{N_c}\Big [(\bar qq)^2+(\bar q\mathrm{i}\gamma^5\vec\tau q)^2 \Big
].
 \label{40}
\end{eqnarray}

Our goal is the investigation of the ground state properties (or phase structure) of the system, described by the Lagrangian (\ref{40}), and its dependence on the chemical potentials $\mu_B$, $\mu_I$ and $\mu_{I5}$ (both at zero and nonzero temperature). It is well known that all information on the phase structure of the model is contained in its thermodynamic potential (TDP). Namely, in the behavior of its global minimum point vs. chemical potentials. In order to find the TDP of the model, we start from a semibosonized version of the Lagrangian (\ref{40}), which contains composite bosonic fields $\sigma (x)$ and $\pi_a (x)$:
\begin{eqnarray}
{\cal L}\ds =\bar q\Big [\gamma^\rho\mathrm{i}\partial_\rho - m_0 +\mu\gamma^0
+ \nu\tau_3\gamma^0+\nu_{5}\tau_3\gamma^0\gamma^5-\sigma
-\mathrm{i}\gamma^5\pi_a\tau_a\Big ]q
 -\frac{N_c}{4G}\Big [\sigma\sigma+\pi_a\pi_a\Big ].
\label{2}
\end{eqnarray}
Here, $a=1,2,3$ and also we introduced the notations $\mu\equiv\mu_B/3$, $\nu\equiv\mu_I/2$ and $\nu_{5}\equiv\mu_{I5}/2$. From the auxiliary Lagrangian (\ref{2}) one gets the equations for the bosonic fields:
\begin{eqnarray}
\sigma(x)=-2\frac G{N_c}(\bar qq);~~~\pi_a (x)=-2\frac G{N_c}(\bar q
\mathrm{i}\gamma^5\tau_a q).
\label{200}
\end{eqnarray}
Note that the composite bosonic field $\pi_3 (x)$ can be identified with the physical $\pi^0(x)$-meson field, whereas the physical $\pi^\pm (x)$-meson fields are the following combinations of the composite fields, $\pi^\pm (x)=(\pi_1 (x)\mp i\pi_2 (x))/\sqrt{2}$.
Obviously, the semibosonized Lagrangian ${\cal L}$ is equivalent to the initial Lagrangian (\ref{40}) when using the equations (\ref{200}).

Starting from the auxiliary Lagrangian (\ref{2}), one obtains in the leading order of the large-$N_c$ expansion (i.e. in the one-fermion loop approximation) the following expression for the effective action ${\cal S}_{\rm {eff}}(\sigma,\pi_a)$ of the bosonic $\sigma (x)$ and $\pi_a (x)$ fields:

\begin{equation}
{\cal S}_{\rm {eff}}(\sigma(x),\pi_a(x))
=-N_c\int
d^4x\left[\frac{\sigma^2(x)+\pi^2_a(x)}{4G}\right]-\mathrm{i}N_c{\rm
Tr}_{sfx}\ln D,
\label{6}
\end{equation}
where the Tr-operation stands for the trace in spinor- ($s$), flavor-($f$) as well as four-dimensional coordinate- ($x$) spaces, respectively. And we have introduced the notation $D$,
\begin{equation}
D\equiv\gamma^\nu\mathrm{i}\partial_\nu -m_0+\mu\gamma^0
+ \nu\tau_3\gamma^0+\nu_{5}\tau_3\gamma^0\gamma^5-\sigma (x) -\mathrm{i}\gamma^5\pi_a(x)\tau_a,
\label{5}
\end{equation}
for the Dirac operator, which acts in the flavor-, spinor- as well as coordinate spaces only.

The ground state expectation values $\vev{\sigma(x)}$ and
$\vev{\pi_a(x)}$ of the composite bosonic fields are determined by
the saddle point equations,
\begin{eqnarray}
\frac{\delta {\cal S}_{\rm {eff}}}{\delta\sigma (x)}=0,~~~~~
\frac{\delta {\cal S}_{\rm {eff}}}{\delta\pi_a (x)}=0,
\label{05}
\end{eqnarray}
where $a=1,2,3$. Just the knowledge of $\vev{\sigma(x)}$ and
$\vev{\pi_a(x)}$ and, especially, of their behaviour vs. chemical potentials supplies us with a phase structure of the model.

In the present work we suppose that in the ground state of the system the quantities $\vev{\sigma(x)}$ and $\vev{\pi_a(x)}$ do not depend on spacetime coordinates $x$,
\begin{eqnarray}
\vev{\sigma(x)}\equiv \sigma,~~~\vev{\pi_a(x)}\equiv \pi_a, \label{8}
\end{eqnarray}
where $\sigma$ and $\pi_a$ ($a=1,2,3$) are already spatially independent constant quantities. In fact, they are coordinates of the global minimum point of the thermodynamic potential (TDP) $\Omega (\sigma,\pi_a)$.
In the leading order of the large-$N_c$ expansion %and using (\ref{8})
it is defined by the following expression:
\begin{equation}
\int d^4x \Omega (\sigma,\pi_a)=-\frac{1}{N_c}{\cal S}_{\rm
{eff}}\big (\sigma(x),\pi_a (x)\big )\Big|_{\sigma
(x)=\sigma,\pi_a(x)=\pi_a} .\label{08}
\end{equation}

In what follows we are going to investigate the $\mu,\nu,\nu_{5}$-dependence of the global minimum point of the function $\Omega (\sigma,\pi_a)$ vs $\sigma,\pi_a$. Let us note that in the chiral limit (due to a $U_{I_3}(1)\times U_{AI_3}(1)$ invariance of the model) the TDP (\ref{08}) depends effectively only on the combinations $\sigma^2+\pi_3^2$ and $\pi_1^2+\pi_2^2$. Whereas at the physical point (i.e. at $m_0\ne 0$ when the relations $\vev{\sigma(x)}\ne 0$ and $\vev{\pi_3(x)} = 0$ are always satisfied) it depends effectively on the combination $\pi_1^2+\pi_2^2$ as well as on $\sigma$ and $\pi_3$. So without loss of generality, in both cases, at $m_0=0$ or $m_0\ne 0$, one can put $\pi_2=\pi_3=0$ in (\ref{08}), and study the TDP as a function of only two variables. To simplify the notations, we introduce the following $M\equiv\sigma+m_0$ and $\Delta\equiv\pi_1$ notations, and throughout the paper use the ansatz
\begin{eqnarray}
\vev{\sigma(x)}=M-m_0,~~~\vev{\pi_1(x)}=\Delta,~~~\vev{\pi_2(x)}=0,~~~ \vev{\pi_3(x)}=0. \label{06}
\end{eqnarray}
If in the global minimum point of the TDP we have $\Delta=0$, then isospin $U_{I_3}(1)$ symmetry of the model is spontaneously broken down. After straightforward and well described calculations  \cite{EK2006,EK2006n,AndersenNJL,Khunjua2017mkc} TDP has the following form:
\begin{eqnarray}
\Omega (M,\Delta)=\frac{(M-m_0)^2+\Delta^2}{4G}-\frac{1}{2\pi^2}\sum_{i=1}^{4}\int_{0}^{\Lambda}p^2\big (|\eta_{i}|+\theta(\mu-|\eta_{i}|)(\mu-|\eta_{i}|)\big )dp,
\label{26}
\end{eqnarray}
where $\Lambda$ is a three-momentum cutoff parameter and $\eta_i$ are the roots of the following polynomial:
\begin{eqnarray}
\label{10}
&&\big(\eta^4-2a\eta^2-b\eta+c\big)\big(\eta^4-2a\eta^2+b\eta+c\big )=0,\\
&&a=M^2+\Delta^2+|\vec p|^2+\nu^2+\nu_{5}^2;\\
&&b=8|\vec p|\nu\nu_{5};\nonumber\\
&&c=a^2-4|\vec p|^2(\nu^2+\nu_5^2)-4M^2\nu^2-4\Delta^2\nu_5^2-4\nu^2\nu_5^2. \nonumber
\end{eqnarray}

It is evident from (\ref{10}) that the TDP (\ref{26}) is an even
function over the variable $\Delta$, and parameters $\nu$ and $\nu_5$. In addition, it is invariant under the transformation $\mu\to-\mu$. Hence, without loss of generality we can consider in the following only $\mu\ge 0$, $\nu\ge 0$, $\nu_5\ge 0$, and $\Delta\ge 0$ values of these quantities. Moreover in the chiral limit, the TDP (\ref{26}) is invariant with respect to the so-called duality transformation ${\cal D}:~M\longleftrightarrow \Delta,~~\nu\longleftrightarrow\nu_5$, which was ultimately investigated in \cite{Khunjua2017mkc,Khunjua:2018sro} and as we will see stay instructive feature even at physical point ($m_0\neq0$).

Using well known thermal summation technique \cite{kapusta2006finite}, one can obtain following expression for the TDP $\Omega_T(M,\Delta)$:
\begin{eqnarray}
\label{TDPT}
\Omega_T (M,\Delta) = \Omega (M,\Delta) -T\sum_{i=1}^{4}\int_{0}^{\Lambda}\frac{p^2dp}{2\pi^2}\Big\{\ln(1+e^{-\frac{1}{T}(|\eta_{i}-\mu|)})+\ln(1+e^{-\frac{1}{T}(|\eta_{i}+\mu|)})\Big\},\label{260}
\end{eqnarray}
where $\Omega (M,\Delta)$ is the TDP (\ref{26}) of the system at zero temperature.

Technically, to define the ground state of the system one should find the coordinates $(M_0,\Delta_0)$ of the global minimum point (GMP) of the TDP (\ref{260}). Since the NJL model is a non-renormalizable theory we have to use fitting parameters for the quantitative investigation of the system. We use the following, widely used parameters:
$$
m_0 = 5,5 \,{\rm MeV};\qquad G=15.03\, {\rm GeV}^{-2};\qquad \Lambda=0.65\, {\rm GeV}.
$$
In this case at $\mu=\nu=\nu_5=0$ one gets for constituent quark mass the value $M=309\,{\rm MeV}$.

As our main goal of the present paper is to prove the possibility of the charged PC phenomenon in hot dense quark matter (at least in the framework of the NJL model (1)), the consideration of the physical quantity $n_{q}$, called quark number density, is now in order. It is related to the baryon number density as $n_{q}=3n_B$ because $\mu=\mu_B/3$. In the general case this quantity is defined by the relation $n_q=-\frac{\partial\Omega(M_0,\Delta_0)}{\partial\mu}$ where $M_0$ and $\Delta_0$ are coordinates of the GMP of a thermodynamic potential.

There are the following phases that could be realized in the system under different external circumstances:
\begin{itemize}
	\item $M=0; \Delta=0$ -- symmetrical phase (it could be realized only in chiral limit $m_0=0$)
	\item $M\ne0; \Delta=0; n_q=0$ -- chiral symmetry breaking phase ({\bf CSB})
	\item $M\ne0; \Delta\ne0; n_q=0$ -- charged pion condensation phase with zero quark density ({\bf PC}) \mbox{($M=0$ in the chiral limit)}.
	\item \mbox{$M\ne0; \Delta=0; n_q\ne0$ -- chiral symmetry breaking phase with nonzero quark density ({\bf CSB${\rm \bf _{d}}$})}.
	\item $M\ne0; \Delta\ne0; n_q\ne0$ -- charged pion condensation phase with nonzero quark density ({\bf PC${\rm \bf _{d}}$}).
\item $M\approx m_0; \Delta=0$ -- approximate symmetrical phase ({\bf ApprSYM}). In this case quark condensate $M$ is of order of the bare quark mass $m_0$, so in the limit $m_0\to 0$ this phase turns into an exactly symmetrical phase with $M=0$.
\end{itemize}
Below we will investigate phase portrait using this definitions. 
%-------------------------------Phase structure of the model-----------
\section{Phase structure of the model}
%-----------------------------------------------------------------------
%------------------------------------m=0--------T\neq0------------------
\subsection{Phase portrait in chiral limit $(m_0=0)$ and zero temperature $(T=0)$}
%-----------------------------------------------------------------------
Let us start from the chiral limit with zero temperature. Although this case has been investigated in the article \cite{Khunjua2017mkc} it is useful to recall main features of the phase portrait in the massless case. There are $(\nu,\nu_5)$-phase portraits of the model in the Fig.1: the left with $\mu=0\,{\rm MeV}$ and the right with $\mu=150\,{\rm MeV}$. These phase diagrams well illustrate the fact that $\nu_5$-chemical potential does promote PC phase with non-zero quark density (PC$_{\rm d}$-phase).

Also, it is easy to see that all PC phases are arranged mirror symmetrically to all CSB phases with respect to the line $\nu=\nu_5$. It is a result of certain duality symmetry ${\cal D}$ of the TDP (\ref{26}), i.e. it is invariant under the ${\cal D}:~M\longleftrightarrow \Delta,~~\nu\longleftrightarrow\nu_5$ transformation which could be strictly seen from (\ref{10}). On the one hand, the duality symmetry helps to investigate the phase portrait
because one needs to calculate only one half of the phase portrait meanwhile another one can be obtained using the symmetry ${\cal D}$. On the other hand, the duality could be a result of some more fundamental symmetry of the Lagrangian. For example, Pauli--G\"ursey symmetry in (1+1)-dimensions leads to the very similar duality between CSB and superconducting phase in the framework of the NJL$_2$ model \cite{Ebert2014woa}. So probably, the duality in (3+1)-dimensions can also be a consequence of some internal symmetry of the model Lagrangian and is inherent property of the dense quark matter.
\begin{figure*}
\centering
\includegraphics[width=0.85\textwidth]{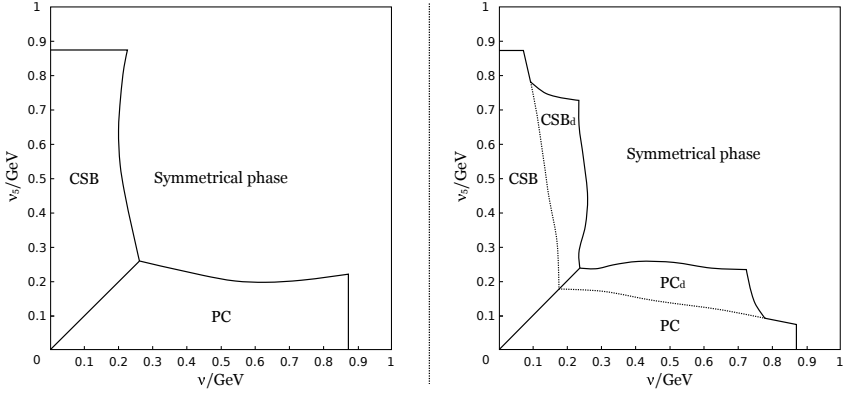}
\caption{The $(\nu,\nu_5)$-phase portrait of the model in the chiral limit $(m_0=0)$ for $\mu=0\,{\rm MeV}$ -- left side and $\mu=150\,{\rm MeV}$ -- right side. Notations are defined in the end of section \ref{sec-1}.}
\label{fig-1}
\end{figure*}
%------------------------------------m\neq0--------T=0------------------
\subsection{Phase portrait at the physical point $(m_0\neq 0)$ and zero temperature $(T=0)$}
%-----------------------------------------------------------------------
\begin{figure*}
\centering
\includegraphics[width=0.85\textwidth]{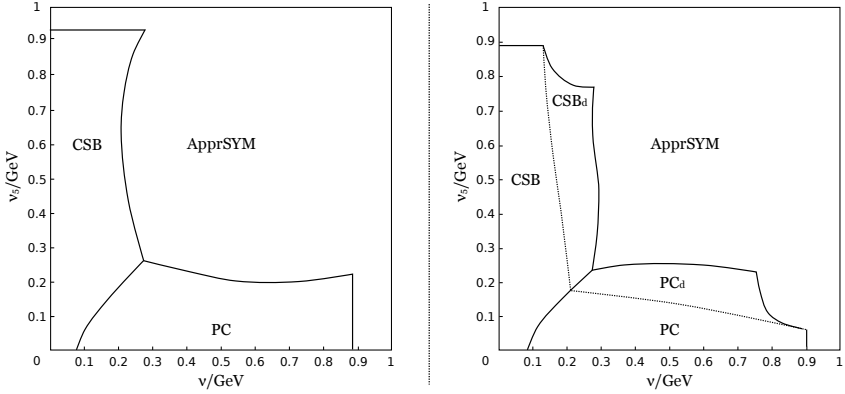}
\caption{The $(\nu,\nu_5)$-phase portrait of the model at the physical point $(m_0=5,5\, {\rm MeV})$ for the same values of the $\mu$ as in Fig.\ref{fig-1}, namely $\mu=0\,{\rm MeV}$ -- left side and $\mu=150\,{\rm MeV}$ -- right side. Notations are defined in the end of section \ref{sec-1}.}
\label{fig-2}
\end{figure*}
Let us move to the main results of the work. $(\nu,\nu_5)$-phase portraits of the model with nonzero bare quark mass
are depicted in Fig.2 for the same values of the quark number chemical potential $\mu$ as in Fig.1: the left with $\mu=0\,{\rm MeV}$ and the right with $\mu=150\,{\rm MeV}$.

First of all, the exact dual symmetry ${\cal D}$ of these phase portraits, that we have observed in the chiral limit, is broken explicitly. Nevertheless duality is still relatively instructive feature even at the physical point. On the other hand, the results become more physically adequate as all known investigations \cite{Son2000xc,Son2001,Splittorff:2000mm,Loewe2002tw}  (including lattice calculations \cite{Gupta,PhysRevD.66.014508, Brandt2017oyy}) predict a threshold $\nu^c=m_\pi/2\approx 70 MeV$ of a second order phase transition to the PC phase, which is presented in Fig.2.
\begin{figure*}
\centering
\includegraphics[width=0.85\textwidth]{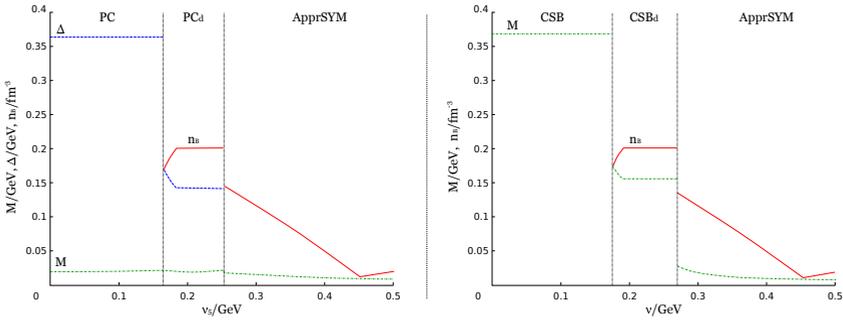}
\caption{The Gaps $M,\Delta$ and baryon density $n_B$ for $\mu=150\,{\rm MeV}$. Left side -- ($M,\Delta, n_B$) vs. $\nu_5$ at $\nu=350\,{\rm MeV}$; Right side -- ($M,\Delta, n_B$) vs. $\nu$ at $\nu_5=350\,{\rm MeV}$. In other words -- slices of the phase portrait on the right side of the Fig.2. }
\label{fig-3}
\end{figure*}
To have more precise picture, take a look at the Fig.3, where the Gaps $M,\Delta$ and baryon density $n_B$ vs. $\nu_5$ and $\nu$ for $\mu=150\,{\rm MeV}$ are depicted, i.e. slices of the phase portrait on the right side of the Fig.2. Left side -- ($M,\Delta, n_B$) vs. $\nu_5$ at $\nu=350\,{\rm MeV}$; Right side -- ($M,\Delta, n_B$) vs. $\nu$ at $\nu_5=350\,{\rm MeV}$ (we choose the same values for $\nu$ and $\nu_5$ to emphasis the dual symmetry ${\cal D}$). It is seen that the quark matter in PC$\rm{_d}$-phase has baryon density approximately equal to the density of the ordinary nuclear matter $n_B$ $\approx0.2$ {\rm fm$^{-3}$}. Also, more importantly, PC${\rm_d}$-phase occupies a rather wide region of $(\nu,\nu_5)$-phase portrait, which confirms that $\nu_{5}$ does promote PC${\rm_d}$-phase even at the physical point.

%------------------------------------m\neq0--------T\neq0------------------
\subsection{Phase portrait at the physical point $(m_0\neq 0)$ and nonzero temperature $(T\neq0)$}
%--------------------------------------------------------------------------
The last thing that we want to discuss is the thermal effects. Though, the effect of non-zero temperatures is quite predictable (one can expect that the temperatures just restore all the broken symmetries of the model), we investigate nonzero temperatures because it is important in a number of applications. We know that compact stars are cold and one can consider their temperatures as zero, but probably there could be scenarios in which the temperatures could be important even in the context of compact stars. So it is instructive to know how robust {\rm PC$\rm{_d}$-phase is under temperature.

Using the formulae (\ref{TDPT}) we calculate the ($T,\nu$)-phase portrait at $\mu=200\,{\rm MeV}, \nu_5=200 \,{\rm MeV}$ (left side of the Fig.4) and its slice with gaps $M,\Delta$ at $\nu=300\,{\rm MeV}$ (right side of the Fig.4). Indeed, broken symmetries are restored with increasing temperature, but it is seen that the PC$\rm{_d}$-phase is quite robust up to $T\approx 50\,{\rm MeV}$.
\begin{figure*}
\centering
\includegraphics[width=0.85\textwidth]{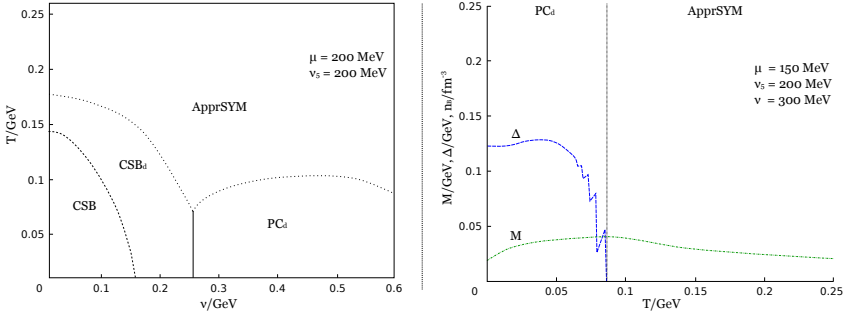}
\caption{The ($T,\nu$)-phase portrait at $\mu=200\,{\rm MeV}, \nu_5=200 \,{\rm MeV}$ (left figure) and it's slice with gaps $M,\Delta$ and baryon density $n_B$ at $\nu=300\,{\rm MeV}$ (right figure). The dotted lines denote second-order phase transition.}
\label{fig-4}
\end{figure*}
%------------------------------------Summary and Conclusions---------------
\section{Summary and Conclusions}
%--------------------------------------------------------------------------
In this work the influence of isotopic and chiral imbalance on phase structure of hot/cold dense quark matter at the physical point ($m_0\neq0$) has been investigated in the framework of the (3+1)-dimensional NJL model with two quark flavors in the large-$N_{c}$ limit ($N_{c}$ is the number of colors). Our special interest was devoted to the charged pion condensation phenomenon and it's existence in the physically adequate circumstances.

It has been shown that chiral imbalance, i.e. difference between left- and right-handed quarks, helps to promote PC${\rm_d}$-phase (PC phase in dense quark matter) in the region of the phase diagram when it is prohibited in the chiral symmetrical case at $\nu_5=0$ (see Fig.1 and Fig.2).

It is well known that nonzero bare quark mass and nonzero temperature tend to suppress PC${\rm_d}$-phase. We have established that the promotion of the PC${\rm_d}$-phase due to nonzero $\nu_5$ is a quite strong effect and it is robust even in the hot dense quark matter at the physical point. Taking into account that imbalance between left- and right-handed quarks is highly anticipated effect, especially in the compact stars, one can expect that PC${\rm_d}$-phase could also exist in the real world.

Moreover, we have also shown that the discrete symmetry (duality) between chiral and charged pion condensations, which is exact in the case of massless quarks, remains a quite good approximation and very instructive feature of the phase diagram even if the quarks have bare mass.

\bibliography{pion_condensation}

\end{document}